\documentclass[11pt, onecolumn]{article}
\pdfoutput = 1
\usepackage{amsmath}
\usepackage{amsfonts}
\usepackage{natbib}
\usepackage{graphicx}
\usepackage{subfigure}

\usepackage{mdwlist}
\usepackage{tabularx}
\usepackage{multirow}
\usepackage{url}
\usepackage[vlined,algoruled,linesnumbered]{algorithm2e}
\usepackage[colorlinks,linkcolor=red,citecolor=blue,urlcolor=blue]{hyperref}
\usepackage{tikz}
\usetikzlibrary{shapes,shadows,positioning,calc,patterns}

\input{Definitions}

 \title{New Approximation Algorithms for Minimum
  Enclosing Convex Shapes} 
\author{
  Ankan Saha\thanks{
  Department of Computer Science
  University of Chicago
  \texttt{ankans@cs.uchicago.edu}} 
  \and
  {\bf S.V$\!.\,$N. Vishwanathan} \thanks{
  Department of Statistics and Computer Science 
  Purdue University 
  \texttt{vishy@stat.purdue.edu}}
  \and 
  {\bf Xinhua Zhang} \thanks{
    Department of Computing Science
    University of Alberta
    \texttt{xinhua.zhang.cs@gmail.com}}
}
\begin{document}
\maketitle

\thispagestyle{empty}

\begin{abstract}
  Given $n$ points in a $d$ dimensional Euclidean space, the Minimum
  Enclosing Ball (MEB) problem is to find the ball with the smallest
  radius which contains all $n$ points. We give a
  $O(nd\Qcal/\sqrt{\epsilon})$ approximation algorithm for producing
  an enclosing ball whose radius is at most $\epsilon$ away from the
  optimum (where $\Qcal$ is an upper bound on the norm of the
  points). This improves existing results using \emph{coresets}, which
  yield a $O(nd/\epsilon)$ greedy algorithm. Finding the Minimum
  Enclosing Convex Polytope (MECP) is a related problem wherein a
  convex polytope of a fixed shape is given and the aim is to find the
  smallest magnification of the polytope which encloses the given
  points. For this problem we present a $O(mnd\Qcal/\epsilon)$
  approximation algorithm, where $m$ is the number of faces of the
  polytope. Our algorithms borrow heavily from convex duality and
  recently developed techniques in non-smooth optimization, and are in
  contrast with existing methods which rely on geometric arguments. In
  particular, we specialize the excessive gap framework of
  \citet{Nesterov05a} to obtain our results.
\end{abstract}
\newpage 

\section{Introduction}
\label{sec:Introduction}
\vspace{-1em}

Given a set $S = \{\xb_{1}, \xb_{2}, \ldots, \xb_{n}\}$ of $n$ points in
$\RR^{d}$, the minimum enclosing ball (MEB) is the ball with the
smallest radius which contains all the points in $S$. The problem of
finding a MEB arises in application areas as diverse as data mining,
learning, statistics, computer graphics, and computational geometry
\citep{ElzHea72}. Therefore efficient algorithms for this problem are
not only of theoretical interest, but also have wide practical
applicability.

Exact algorithms for finding the MEB typically have an exponential
dependence on $d$ \citep{Meg84, Welzl91}. For example, the
\citet{Welzl91} algorithm runs in $O(n (d+1) (d+1)!)$ time which makes
it inadmissible for many practical applications; in the case of linear
SVMs data may have a million or more dimensions. Therefore, there has
been a significant interest in finding approximation algorithms for this
problem.

State of the art approximation algorithms for the MEB problem
extensively use the concept of coresets
\citep{Clarkson08,BadCla02,Panigrahy04,Yildirim08}. Given an $\epsilon >
0$, an $\epsilon$-coreset $S' \subset S$ has the property that if the
smallest enclosing ball containing $S'$ is expanded by a factor of $(1 +
\epsilon)$, then the resulting ball also contains $S$. Therefore,
locating an $\epsilon$-coreset is equivalent to finding an $(1+
\epsilon)$ approximation algorithm to the MEB problem. The approximation
guarantees of such algorithms are \textbf{multiplicative}. Briefly, a
coreset is built incrementally in a greedy fashion
\citep{Clarkson08}. At every iteration, the MEB of the current candidate
coreset is built. If every point in $S$ lies in an $(1+\epsilon)$ ball
of the current solution then the algorithm stops, otherwise the most
\emph{violated} point, that is, the point which is furthest away from
the current MEB is included in the candidate coreset and the iterations
continue. The best known algorithms in this family have a running time
of $O(nd/\epsilon)$ \citep{Panigrahy04, Clarkson08}.

In contrast, we present a new algorithm which is derived by casting
the problem of finding the MEB as a convex but non-smooth optimization
problem. By specializing a general framework of \citet{Nesterov05a},
our algorithm is able to achieve a running time of
$O(nd\Qcal/\sqrt{\epsilon})$ where $\Qcal$ is an upper bound on the
norm of the points. Also, the approximation guarantees of our
algorithm are \textbf{additive}, that is, given a tolerance $\epsilon
>0$ and denoting the optimal radius by $R^{*}$, our algorithm produces
a function whose value lies between ${R^{*}}^{2}$ and
${R^{*}}^{2}+\epsilon$. Although these two types of approximation
guarantees seem different, by a simple argument in section
\ref{sec:MultversAddit}, we show that our algorithm also yields a
traditional scale-invariant $\epsilon$ multiplicative approximation
with $O(nd\Qcal/\sqrt{\epsilon})$ effort.

We extend our analysis to the closely related minimum enclosing convex
polytope (MECP) problem, and present a new algorithm. As before, given a
set $S = \{\xb_{1}, \xb_{2}, \ldots, \xb_{n}\}$ of $n$ points in
$\RR^{d}$, the task here is to find the smallest polytope of a given
fixed shape which encloses these points. In our setting translations and
magnifications are allowed but rotations are not allowed. 
We present a $O(mnd\Qcal/\epsilon)$ approximation algorithm, where $m$
denotes the number of faces of the polytope. 

We apply our algorithms to two problems of interest in machine learning
namely finding the maximum margin hyperplane and computing the distance
of a polytope from the origin. A coreset algorithm for the first problem
was proposed by \citet{HarRotZim07} while the second one was studied by
\citet{GaeJag09}. In both cases our algorithms require fewer number of
iterations and yield better computational complexity bounds. 


Our paper is structured as follows: In Section \ref{sec:defn} we
introduce notation, briefly review some results from convex duality, and
present the general framework of \citet{Nesterov05a}. In Section
\ref{sec:MEB} we address the MEB problem and in Section \ref{sec:MECP}
the MECP problem, and present our algorithms and their analysis. We
discuss some applications of our results to machine learning problems in
Section \ref{sec:application}.
The paper then concludes with a discussion and outlook for the future
in Section \ref{sec:Conclusions}. Technical proofs can be found in
Appendix \ref{sec:app_proof_exc_cond_meet} and
\ref{sec:simple_qp}, while preliminary experimental evaluation can be
found in Appendix~\ref{sec:exp}. 


\section{Definitions and Preliminaries}
\label{sec:defn}

In this paper, lower bold case letters (\eg, $\wb$, $\mub$) denote
vectors, while upper bold case letters (\eg, $\Ab$) denote matrices or
linear operators. We use $w_{i}$ to denote the $i$-th component of
$\wb$, $A_{ij}$ to denote the $(i, j)$-th entry of $\Ab$, and
$\inner{\wb}{\wb'} := \sum_{i} w_{i} w'_{i}$ to denote the Euclidean dot
product between vectors $\wb$ and $\wb'$.  $\Delta_k$ denotes the $k$
dimensional simplex. Unless specified otherwise, $\nbr{\cdot}$ refers to
the Euclidean norm $\|\wb\| := \sqrt{\inner{\wb}{\wb}} = \left(
  \sum_{i=1}^{n} w_{i}^{2}\right)^{\frac{1}{2}}$. For a matrix $\Ab \in
\RR^{n \times d}$, we have the following definition of the norm
\begin{align*}
  \|\Ab\| = \max\cbr{\inner{\Ab\wb}{\ub}: \|\wb\| = 1, \|\ub\| = 1}.
\end{align*}
We also denote $\overline{\RR} := \RR \cup \{\infty\}$, and $[t]:= \{1,
\ldots, t\}$.

\begin{definition}
  \label{def:eps-accurate-sol}
  Let $Q_1 \subseteq \RR^n$, $f: Q_1 \to \RRbar$, and $f^{*} :=
  \min_{\wb \in Q_{1}} f(\wb) < \infty$. A point $\wb' \in Q_{1}$ such
  that
  \begin{align}
    \label{eq:eps-accurate-sol}
    f(\wb') \leq f^{*} + \epsilon
  \end{align}
  is said to be an $\epsilon$-accurate minimizer of $f$. We will also
  sometimes call $\wb'$ an $\epsilon$-accurate solution.
\end{definition}
The following three standard concepts from convex analysis (see \eg\
\citet{HirLem93}) are extensively used in the sequel.
\begin{definition}
  \label{def:strong-convex}
  A convex function $f:\RR^{n} \to \RRbar$ is strongly
  convex with respect to a norm $\|\cdot\|$ if there exists a constant
  $\rho > 0$ such that $f - \frac{\rho}{2} \|\cdot\|^{2}$ is convex.
  $\rho$ is called the modulus of strong convexity of $f$, and for
  brevity we will call $f$ $\rho$-strongly convex or $\rho$-s.c.
\end{definition}
\begin{definition}
  \label{def:lip-cont-grad}
  Suppose a function $f: \RR^n \to \RRbar$ is differentiable on $Q \subseteq
  \RR^n$.  Then $f$ is said to have Lipschitz continuous gradient (\lcg) with
  respect to a norm $\|\cdot\|$ if there exists a constant $L$ such that
  \begin{align}
    \label{eq:lip-cont-grad}
    \| \nabla f(\wb) - \nabla f(\wb')\| \leq L \| \wb - \wb'\| \qquad
    \forall\ \wb, \wb'\in Q.
  \end{align}
For brevity, we will call $f$ $L$-\lcg.
\end{definition}
\begin{definition}
  \label{def:fenchel_dual}
  The Fenchel dual of a function $f: \RR^n \to \RRbar$ is a function $f^{\star}:
  \RR^n \to \RRbar$ defined by
  \begin{align}
    \label{eq:fenchel-dual}
    f^{\star}(\wb^{\star}) = \sup_{\wb \in \RR^n}
    \cbr{\inner{\wb}{\wb^{\star}} - f(\wb)}
  \end{align}
\end{definition}
Strong convexity and Lipschitz continuity of the gradient are related by
Fenchel duality according to the following lemma:
\begin{lemma}[{\citet[][Theorem 4.2.1 and 4.2.2]{HirLem93}}]
$\phantom{.}$
\label{theorem:SC_LCG}
\begin{enumerate}
\item If $f: \RR^n \to \RRbar$ is $\rho$-s.c., then $f^{\star}$ is
  finite on $\RR^n$ and $f^{\star}$ is $\frac{1}{\rho}$-\lcg.
\item If $f: \RR^n \to \RR$ is convex, differentiable on $\RR^n$, and $L$-\lcg,
  then $f^{\star}$ is $\frac{1}{L}$-s.c.
\end{enumerate}
\end{lemma}

\subsection{Nesterov's Framework}
\label{sec:NesterovsFramework}

In sections \ref{sec:MEB} and \ref{sec:MECP} we will show that the MEB
and MECP problems respectively can be cast as minimizing convex
non-smooth objective functions. In a series of papers, Nesterov
\citep{Nesterov83, Nesterov05, Nesterov05a} proposed a general
framework for this task, which we now briefly review.

Let $Q_{1}$ and $Q_{2}$ be subsets of Euclidean spaces and $\Ab$ be a
linear map from $Q_1$ to $Q_2$.  Suppose $f$ and $g$ are convex
functions defined on $Q_1$ and $Q_2$ respectively, and we are interested
in the following optimization problem:
\begin{align}
  \min_{\wb \in Q_{1}} J(\wb) 
\text{ where }
  J(\wb) := f(\wb) +
  \gstar(\Ab\wb) = f(\wb) + \max_{\ub \in Q_2} \cbr{\inner{\Ab
      \wb}{\ub}-g(\ub)}.
  \label{eq:primal}
\end{align}
We will make the following standard assumptions: a) $Q_2$ is compact; b)
with respect to a certain norm on $Q_1$, the function $f$ defined on
$Q_1$ is $\rho$-s.c.\ but not necessarily \lcg, and c) with respect to a
certain norm on $Q_2$, the function $g$ defined on $Q_{2}$ is
$L_{g}$-\lcg\ and convex, but not necessarily strongly convex.  

\begin{figure}[tbp]
  \centering
  \begin{tabular}[htbp]{ccc}
    \begin{tikzpicture}[domain=-3:3,samples=100,scale=0.6]
      \draw[color=blue,thick] (-3, 3) -- (-1.7, 1) -- (-0.75, 0.2) --
      (0, 0) -- (0.75, 0.2) -- (1.7, 1.0) -- (3.0, 3.0) node[right]
      {$J(\wb)$}; 
      
      \draw[color=black,thick] plot (\x,-0.1*\x^2) node[right] {$D(\ub)$}; 

      \node[fill,draw,star, inner sep=0.8pt,color=red,thick] at (0, 0) {} ;

    \end{tikzpicture} & 
    \begin{tikzpicture}[domain=-3:3,samples=100,scale=0.6]
      \draw[color=blue,thick]    (-3, 3) -- (-1.7, 1) -- (-0.75, 0.2) -- (0, 0)
      -- (0.75, 0.2) -- (1.7, 1.0) -- (3.0, 3.0) node[right] {$J(\wb)$};
      
      \draw[color=black,thick] plot (\x,-0.1*\x^2) node[right] {$D(\ub)$};
      
      \draw[color=blue,thick] plot (\x,0.25*\x^2-0.7) node[right] {$J_{\mu_{k}}(\wb)$};; 
      
      \draw[<->, thick, color=red] (-2.5, 0.8625) -- (-1.8, 1.2125) node[right,color=black]
      {$\mu_{k} \Dcal$};
      
      \node[fill,draw,star, inner sep=0.8pt,color=red,thick] at (0, 0) {} ;
      
    \end{tikzpicture} &
    \begin{tikzpicture}[domain=-3:3,samples=100,scale=0.6]
      \draw[color=blue,thick]    (-3, 3) -- (-1.7, 1) -- (-0.75, 0.2) -- (0, 0)
      -- (0.75, 0.2) -- (1.7, 1.0) -- (3.0, 3.0) node[right] {$J(\wb)$};
      
      \draw[color=black,thick] plot (\x,-0.1*\x^2) node[right] {$D(\ub)$};
      
      \draw[color=blue,thick] plot (\x,0.25*\x^2-0.7) node[right] {$J_{\mu_{k}}(\wb)$};; 

      \draw[<->, thick, color=red] (-2.5, 0.8625) -- (-1.8, 1.2125) node[right,color=black]
      {$\mu_{k} \Dcal$};
      
      \node[fill,draw,circle, inner sep=1pt,color=blue,thick] at (-1, -0.45) {} ;
      
      \node[fill,draw,star, inner sep=0.8pt,color=red,thick] at (0, 0) {} ;
      
      \draw node [fill,draw,circle, inner sep=1pt,color=black,thick] at
      (1, -0.1) {};
      
      \node[] at (-1.5, -1.8) {$J_{\mu_{k}}(\wb_{k})$};
      
      \draw[->, thick] (-1.7, -1.4) -- (-1, -0.6);
    
      \node[] at (0.5, 2.2) {$D(\ub_{k})$};
      
      \draw[->, thick] (0.5, 1.8) -- (1, 0);
    
    \end{tikzpicture}    
  \end{tabular}
  \caption{Illustration of excessive gap.  By strong convexity, the dual
    $D(\ub)$ is always a lower bound to the primal $J(\wb)$ (left). We
    approximate $J(\wb)$ by a smooth lower bound $J_{\mu_{k}}(\wb)$ (middle).
    Since $D(\ub_{k})$ is sandwiched between $J_{\mu_{k}}(\wb)$ and
    $J(\wb)$, as $\mu_{k} \to 0$ we get closer and closer to the true
    optimum (right). 
}
    \label{fig:ex_gap}
\end{figure}
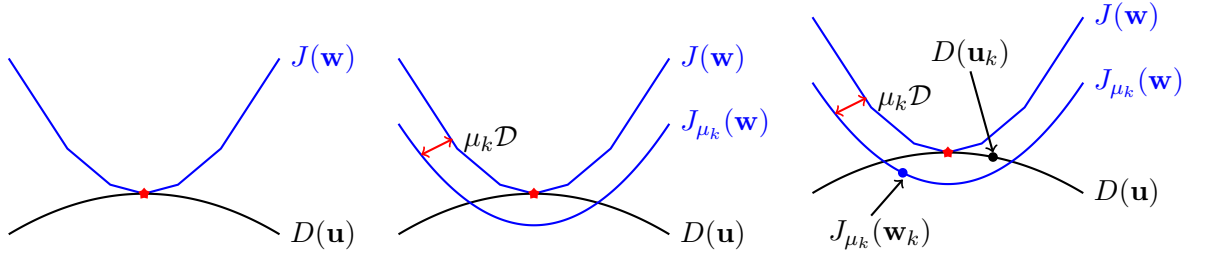

The key difficulty in solving \eqref{eq:primal} arises because
$\gstar$ and hence $J$ may be non-smooth. Our aim is to uniformly
approximate $J(\wb)$ with a smooth and strongly convex function.
Towards this end let $d$ be a $\sigma$-s.c.\ smooth function with the
following properties:
\begin{align*}
  \min_{\ub \in Q_2} d(\ub) = 0,
  \quad \ub_{0} = \argmin_{\ub \in Q_2} d(\ub),
  \text{ and }  \Dcal := \max_{\ub \in Q_2} d(\ub).
\end{align*}
In optimization parlance $d$ is called a prox-function. For a positive
constant $\mu \in \RR$ define
\begin{align}
  \label{eq:reg_primal}
  J_{\mu}(\wb) &:= 
  f(\wb) + \max_{\ub \in Q_2} \cbr{\inner{\Ab \wb}{\ub}-g(\ub) - \mu \,
    d(\ub)}.
\end{align}
It can be easily verified that $J_{\mu}$ is not only smooth and convex
but also $\frac{1}{\sigma \mu} \|\Ab\|^{2}$-\lcg\, 
\citep{Nesterov05}. Furthermore, if $\Dcal < \infty$ then $J_{\mu}$ is
uniformly close to $J$, that is,
\begin{align}
  \label{eq:approx_relation}
  J_\mu(\wb)\leq J(\wb)\leq J_\mu(\wb)+ \mu \, \Dcal \ .
\end{align}
If some mild constraint qualifications hold \citep[\eg\ Theorem
3.3.5][]{BorLew00} one can write the dual $D(\ub)$ of $J(\wb)$ using
$\Ab^{\top}$ (the transpose of $\Ab$) as
\begin{align}
  \label{eq:dual}
  D(\ub) := -g(\ub) - \fstar(-\Ab^{\top} \ub) = -g(\ub) - \max_{\wb
    \in Q_{1}} \cbr{\inner{-\Ab \wb}{\ub} - f(\wb)},
\end{align}
and assert the following:
\begin{align}
  \label{eq:borlew}
  \inf_{\wb \in Q_{1}} J(\wb) = \sup_{\ub \in Q_{2}} D(\ub), \quad
  \text{and} \quad J(\wb) \ge D(\ub) \quad \forall\ \wb \in Q_{1}, \ub
  \in Q_{2}.
\end{align}
The key idea of excessive gap minimization pioneered by
\citet{Nesterov05a} is to maintain two estimation sequences
$\cbr{\wb_{k}}$ and $\cbr{\ub_{k}}$, together with a diminishing
sequence $\cbr{\mu_{k}}$ such that
\begin{equation}
  \label{eq:excessive_gap_condition}
  \boxed{J_{\mu_{k}}(\wb_{k}) \le D (\ub_{k}), \text{ and } \lim_{k \to \infty} \mu_{k} = 0.}
\end{equation}
The idea is illustrated in Figure \ref{fig:ex_gap}.  In conjunction
with \eqref{eq:borlew} and \eqref{eq:approx_relation}, it is not hard
to see that $\cbr{\wb_{k}}$ and $\cbr{\ub_{k}}$ approach the solution
of $\min_{\wb} J(\wb) = \max_{\ub} D(\ub)$. Using
\eqref{eq:approx_relation}, \eqref{eq:reg_primal}, and
\eqref{eq:excessive_gap_condition}, we can derive the following bound
on the duality gap:
\begin{align}
  \label{eq:roc}
  J(\wb_{k}) - D(\ub_{k}) \le J_{\mu_{k}}(\wb_{k}) + \mu_{k} \Dcal - D
  (\ub_{k}) \le  \mu_{k} \Dcal.
\end{align}
In other words, the duality gap is reduced at the same rate at which
$\mu_{k}$ approaches $0$. To turn this idea into an implementable
algorithm we need to answer the following two questions:
\begin{enumerate}
\item How to efficiently find initial points $\wb_1$, $\ub_{1}$ and
  $\mu_{1}$ that satisfy \eqref{eq:excessive_gap_condition}.
\item Given $\wb_{k}$, $\ub_{k}$, and $\mu_{k}$, how to
  \emph{efficiently} find iterates $\wb_{k+1}$, $\ub_{k+1}$, and
  $\mu_{k+1}$ which maintain \eqref{eq:excessive_gap_condition}. To
  achieve the best possible convergence rate it is desirable to anneal
  $\mu_{k}$ as fast as possible while still allowing $\wb_{k}$ and
  $\ub_{k}$ to be updated efficiently.
\end{enumerate}
We will now show how the MEB and MECP problems can be cast as convex
optimization problems and derive implementable algorithms by
answering the above questions.

\section{Minimum Enclosing Ball}
\label{sec:MEB}

Given a set of $n$ points $S = \cbr{\xb_1, \ldots, \xb_{n}}$ in a $d$
dimensional space $\RR^d$, a Euclidean ball $B(\cbb, R)$ of radius $R$
centered at $\cbb$ is said to be an enclosing ball if $\forall i \in
[n]$, $\xb_{i} \in B(\cbb, R)$. 

\subsection{Formulation as an Optimization Problem}
\label{sec:Formulationasan}

Clearly, $\xb_{i} \in B(\cbb, R)$ if, and only if, $\|\cbb -
\xb_{i}\|^{2} \leq R^{2}$. Using this observation, the MEB problem can
be cast as the following optimization problem:
\begin{align*}
  \min_{R \in \RR} R \qquad 
  \text{s.t. } \|\cbb-\xb_i\|^{2} \leq R^2 \quad \forall i, 
\end{align*}
which in turn can be reformulated as
\begin{align}
  \label{eq:MEB}
  \min_{\cbb\in \RR^d}\max_{\xb_i\in S} \| \cbb-\xb_i\|^2. 
\end{align}
Rearranging terms
\begin{align}
  \min_{\cbb\in \RR^d}J(\cbb) = \| \cbb\|^2 + \max_{\xb_i \in
    S}\cbr{-2\inner{\cbb}{\xb_i} + \| \xb_i\|^2} 
  = \| \cbb\|^2 + \max_{\ub \in
    \Delta_n}\cbr{\inner{\Ab\cbb}{\ub} + \inner{\ub}{\bb}},
  \label{eq:MEB-primal}
\end{align}
where $\Ab = -2[\xb_1,\xb_2\hdots \xb_n]^{\top}$, and $b_{i} =
\|\xb_i\|^{2}$. Clearly, $J(\cbb)$ can be identified with
\eqref{eq:primal} by setting $Q_{1} = \RR^{d}$, $Q_{2} = \Delta_{n}$,
$f(\cbb) = \nbr{\cbb}^{2}$, and $g(\ub) = - \inner{\ub}{\bb}$. It can be
verified that $g$ is $0$-\lcg, while $f$ is 2-s.c. Therefore, one can
employ Nesterov's framework (Section \ref{sec:NesterovsFramework}) to
minimize $J(\cbb)$. However, as we stated before, we need to specialize
the framework to our setting to obtain an efficient and implementable
algorithm. Towards this end note that the Fenchel dual of
$\nbr{\cdot}^{2}$ is $\frac{1}{4} \nbr{\cdot}^{2}$, and use
\eqref{eq:dual} to write the dual of \eqref{eq:MEB-primal} as
\begin{align}
  \label{eq:MEB-dual}
  D(\ub) = \inner{\ub}{\bb} - \frac{1}{4} \ub^{\top} \Ab \Ab^{\top} \ub.
\end{align}
By using the Cauchy-Schwartz inequality, the gradient 
\begin{align}
  \label{eq:MEB-dual-gradient}
  \nabla D(\ub)  = \bb - \frac{1}{2} \Ab \Ab^{\top} \ub,
\end{align}
can be shown to satisfy 
\begin{align}
  \label{eq:MEB-grad-lipsch}
  \nbr{ \nabla D(\ub_{1}) - \nabla D(\ub_{2})} = \nbr{- \frac{1}{2}
    \Ab \Ab^{\top} \ub_{1} + \frac{1}{2} \Ab \Ab^{\top} \ub_{2}} \leq 
  \frac{1}{2} \nbr{\Ab \Ab^{\top}} \nbr{\ub_{1} - \ub_{2}},
\end{align}
thus establishing that $D(\ub)$ is $\frac{1}{2} \nbr{\Ab
  \Ab^{\top}}$-\lcg. We define $L = \frac{1}{2} \nbr{\Ab
  \Ab^{\top}}$.

Next we turn our attention to the prox-function. Recall that
Nesterov's framework requires a $\sigma$-s.c.\ prox-function on $Q_2 =
\Delta_{n}$; in our case we set
\[d(\ub) = \frac{\sigma}{2} \nbr{\ub - \ub_{0}}^{2}
\]
where $\ub_{0} = \rbr{\frac{1}{n}, \ldots, \frac{1}{n}} \in \RR^n$. For this
choice $\ub_{0} = \argmin_{\ub \in \Delta_{n}} d(\ub)$, $d\rbr{\ub_0} = 0$,
and
\begin{align}
  \label{eq:dcal-bound}
  \Dcal = \max_{\ub \in \Delta_n}d(\ub) = \frac{\sigma}{2} \max_{\ub \in
    \Delta_{n}} \nbr{\ub - \ub_{0}}^{2} = \frac{\sigma}{2}\rbr{1 - \frac{1}{n}}^2 \leq \frac{\sigma}{2}.
\end{align}
Furthermore, for notational convenience, define the following three
maps:
\begin{align}
  \label{eq:c_min}
  \cbb(\ub) &= \argmin_{\cbb \in
    \RR^{d}}\cbr{\inner{\Ab\cbb}{\ub}+\nbr{\cbb}^{2}} \\
  \ub_\mu(\cbb) & = \argmax_{\ub \in
    \Delta_{n}}\cbr{\inner{\Ab\cbb}{\ub} + \inner{\ub}{\bb} - \mu
    d(\ub)} 
  = \argmin_{\ub \in \Delta_{n}}\cbr{\frac{\mu \sigma}{2}
    \nbr{\ub - \ub_{0}}^{2} - \inner{\Ab\cbb + \bb}{\ub}}. 
    \label{eq:u_max} 
  \\
  V(\ub) &= \argmin_{\vb \in
    \Delta_{n}}\cbr{\frac{L}{2}\|\vb-\ub\|^{2}
    - \inner{\grad D(\ub)} {\vb-\ub}} 
  = \argmin_{\vb \in \Delta_{n}}\cbr{\frac{L}{2}\|\vb-\ub\|^{2} -
    \inner{\bb - \frac{1}{2} \Ab \Ab^{\top} \ub} {\vb-\ub}}.
  \label{eq:adj_gradient}
\end{align}
With this notation in place we now describe our excessive gap
minimization method in Algorithm~\ref{algo:MEB}. Unrolling the recursive
update for $\mu_{k}$ yields
\begin{align}
  \label{eq:mu2-update}
  \mu_{k} = (1 - \tau_{k-1})\, \mu_{k-1} = \frac{k}{k+2} \,
  \mu_{k-1} = \frac{(k)(k-1) \ldots 2}{(k+2)(k+1)\ldots 4}
  \frac{L}{\sigma} = \frac{6}{(k+1)(k+2)} \frac{L}{\sigma}.
\end{align}
Plugging this into \eqref{eq:roc} and using \eqref{eq:dcal-bound}
immediately yields the following theorem:
\begin{theorem}[Duality gap]
\label{thm:rate_conv_duality_gap_MEB}
The sequences $\cbr{\cbb_k}$ and $\cbr{\ub_k}$ in Algorithm
\ref{algo:MEB} satisfy
\begin{align}
  \label{eq:convergence}
  J(\cbb_{k}) - D(\ub_{k}) &\le \frac{6L\Dcal}{\sigma (k+1) (k+2)} \leq
  \frac{3L}{(k+1) (k+2)}.
\end{align}
\end{theorem}
As is standard \citep[see \eg][]{Yildirim08}, if we assume that the
input data points lie inside a ball of radius $\Qcal$, that is,
$\max_{i} \nbr{\xb_{i}}\leq \Qcal$ then we can write 
\begin{align}
  \label{eq:L-bound}
  L = \frac{1}{2} \nbr{\Ab \Ab^{\top}} = \frac{1}{2} \max_{\nbr{\cbb} =
    \nbr{\ub} = 1} \cbb^{\top} \Ab \Ab^{\top} \ub = \frac{1}{2}
  \max_{\nbr{\ub} = 1} \nbr{\Ab^{\top} \ub}^{2} = 2 \max_{i}
  \nbr{\xb_{i}}^{2} = 2 \Qcal^{2}.
\end{align}
The last equality follows because $\Ab = -2[\xb_1,\xb_2\hdots
\xb_n]^{\top}$ and the maximum is attained by setting $\ub = \eb_{j}$
where $j = \argmax_{i} \nbr{\xb_{i}}^{2}$. Note that this is just a
conservative estimate and a larger value of $L$ also guarantees
convergence of the algorithm. Plugging this back into
\eqref{eq:convergence} yields
\begin{align}
  \label{eq:convergence-tighter}
  J(\cbb_{k}) - D(\ub_{k}) &\le \frac{6\Qcal^{2}}{(k+1) (k+2)}.
\end{align}
Therefore, to obtain an $\epsilon$ accurate solution of
\eqref{eq:MEB-primal} it suffices to ensure that 
\begin{align}
  \label{eq:convergence-epsilon}
 \frac{6\Qcal^{2}}{(k+1) (k+2)} \leq \epsilon.
\end{align}
Solving for $k$ yields the $O(\Qcal/\sqrt{\epsilon})$ bounds on the number
of iterations as claimed. All that remains is to show that
\begin{theorem}
  \label{thm:excessive_condi_satisfy}
  The update rule of Algorithm \ref{algo:MEB} guarantees that
  \eqref{eq:excessive_gap_condition} is satisfied for all $k \ge 1$.
\end{theorem}
\begin{proof}
  See Appendix  \ref{sec:app_proof_exc_cond_meet}.
\end{proof}

\begin{algorithm}[t]
  \caption{Excessive gap minimization applied to MEB}
  \label{algo:MEB}
  \KwOut{Sequences $\cbr{\cbb_{k}}$, $\cbr{\ub_{k}}$, and $\cbr{\mu_{k}}$
    that satisfy \eqref{eq:excessive_gap_condition}, with
    $\lim_{k \to \infty} \mu_{k} = 0$.}
  Initialize: {Let $\ub_0 = \rbr{\frac{1}{n}, \ldots, \frac{1}{n}}$,
    $\mu_{1} = \frac{L}{\sigma}$, $\cbb_1 = \cbb(\ub_{0})$, $\ub_{1} =
    V\rbr{\ub_{0}}$.}\;

  \For{$k = 1, 2, \ldots$}{

    $\tau_{k} \leftarrow \frac{2}{k+3}$.

    $\betab_k \leftarrow (1 - \tau_k) \ub_k + \tau_k \ub_{\mu_k} (\cbb_k)$.
 
    $\cbb_{k+1} \leftarrow (1 - \tau_k) \cbb_k + \tau_k \cbb(\betab_k)$.

    $\ub_{k+1} \leftarrow V(\betab_k)$.

    $\mu_{k+1} \leftarrow (1 - \tau_{k}) \mu_{k}$.
  }
\end{algorithm}

Each iteration of Algorithm \ref{algo:MEB} requires us to compute
$\cbb(\ub)$, $\ub_{\mu}(\cbb)$, and $V(\ub)$ (see \eqref{eq:c_min},
\eqref{eq:u_max}, and \eqref{eq:adj_gradient}). All other operations
either require constant or linear time. We now show that each of these
three maps can be computed in $O(nd)$ time. This in conjunction with
Theorem \ref{thm:rate_conv_duality_gap_MEB} shows that the time complexity
of our algorithm to find an $\epsilon$ accurate solution of
\eqref{eq:MEB-primal} is $O(nd/\sqrt{\epsilon})$.

By computing the gradient of $\inner{\Ab\cbb}{\ub}+\nbr{\cbb}^{2}$ and
setting it to zero we can show that
\begin{align}
  \label{eq:c-min-cform}
  \cbb(\ub) = -\frac{1}{2} \Ab^{\top} \ub. 
\end{align}
Since $\Ab$ is a $n \times d$ matrix computing $\cbb(\ub)$ takes $O(nd)$
time. On the other hand, computation of $\ub_{\mu}(\cbb)$ can be cast as
the following Quadratic programming (QP) problem with linear constraints:
\begin{align}
  \label{eq:ub-min-qp}
  \min_{\ub}\frac{\mu \sigma}{2} \nbr{\ub}^{2} -
  \inner{\Ab\cbb + \bb + \mu \sigma \ub_{0}}{\ub} \\ 
  \notag
  \text{  s.t. }  \sum_{i} u_{i} = 1 
  \text{ and } 0 \leq u_{i} \leq 1.
\end{align}  
Computing $\Ab\cbb + \bb + \mu \sigma \ub_{0}$ requires $O(nd)$
time. Given $\Ab\cbb + \bb + \mu \sigma \ub_{0}$, we show in Appendix
\ref{sec:simple_qp} that the above QP can be solved in $O(n)$
time. Finally, after some simple algebraic manipulation computation of
$V(\ub)$ can be also be cast as a Quadratic programming (QP) problem
with linear constraints as follows:
\begin{align}
  \label{eq:adj-map-min-qp}
  \min_{\vb} \frac{L}{2}\nbr{\vb}^{2} -
  \inner{L \ub  - \frac{1}{2} \Ab \Ab^{\top} \ub + \bb} {\vb} \\
  \notag
  \text{ s.t. } \sum_{i} v_{i} = 1 
  \text{ and } 0 \leq v_{i} \leq 1.
  \end{align}  
Again, the computational bottleneck is in computing $L \ub - \frac{1}{2}
\Ab \Ab^{\top} \ub + \bb$ which takes $O(nd)$ effort\footnote{To compute
  $\Ab \Ab^{\top} \ub$ efficiently we first compute $\ab = \Ab^{\top}
  \ub$ and then compute $\Ab \ab$.}. Given $L \ub - \frac{1}{2} \Ab
\Ab^{\top} \ub + \bb$ the algorithm in Appendix \ref{sec:simple_qp} can
be applied to solve \eqref{eq:adj-map-min-qp} in $O(n)$ time. 


\subsection{Multiplicative versus Additive Approximation: Scale Invariance}
\label{sec:MultversAddit}

Existing approximation algorithms for the MEB problem based on coresets
provide a multiplicative approximation. Given a set of points, the
coreset algorithms output a center $\cbb$ and radius $R$ such that all
the given points lie inside the ball of radius $R(1 + \epsilon)$
centered at $\cbb$. In contrast, our algorithm produces a center and a
radius $R$ such that ${R^*}^2 \leq R^{2} \leq {R^*}^2 + \epsilon'$, where
$R^{*}$ denotes the radius of the optimal minimum enclosing ball. 

At first glance the two types of guarantees do not look directly
comparable since the additive guarantees seem to vary with change of
scale. To produce an $\epsilon_{M}$ \emph{scale invariant}
multiplicative approximation with our algorithm, set $\epsilon' =
\epsilon_{M} {R^*}^{2}$. In view of \eqref{eq:convergence-epsilon} it
follows that ${R^*}^2 \leq R^{2} \leq {R^*}^2 (1 + \epsilon_{M})$
whenever\footnote{We prove our bounds in terms of $R^2$ but it is
  trivial to convert this to a bound in terms of $R$.}
\begin{align}
  \label{eq:additive-mult}
  \frac{6\Qcal^{2}}{(k+1) (k+2)} \leq {R^{*}}^{2} \epsilon_{M}.
\end{align}
Solving for $k$ obtains 
\begin{align}
  \label{eq:additive-mult-ink}
  \frac{\Qcal}{R^*} \sqrt{\frac{6}{\epsilon_{M}}} \leq k.
\end{align}
However, since $R^*$ is unknown this bound cannot be used as a practical
stopping criterion. Instead, one can use the following observation:
select a arbitrary pair of points $\xb_{i}$ and $\xb_{j}$ from $S$ and
compute $\frac{1}{2} \nbr{\xb_{i} - \xb_{j}}$. Denote this quantity by
$\Pcal$, and let $\cbb^*$ be the center of the optimal MEB. Clearly
\begin{align}
  \label{eq:approx_P}
  \Pcal = \frac{1}{2} \nbr{\xb_{i} - \xb_{j}} \leq \frac{1}{2}
  \nbr{\xb_{i} - \cbb^*} + \frac{1}{2} \nbr{\xb_{j} - \cbb^*} =
  \frac{1}{2}R^{*} + \frac{1}{2} R^{*} = R^{*}.
\end{align}
Therefore replacing $R^{*}$ by $\Pcal$ in \eqref{eq:additive-mult-ink}
yields the following computable upper bound on the number of iterations:
\begin{align}
  \label{eq:additive-mult-practical}
  \frac{\Qcal}{\Pcal} \sqrt{\frac{6}{\epsilon_{M}}} \leq k.
\end{align}
This shows that $O(1/\sqrt{\epsilon_{M}})$ iterations of our algorithm
suffice to produce a $\epsilon_{M}$-multiplicative approximation. 

Note that just like in the case of coreset based algorithms this bound
is scaling invariant, that is, if all the $\xb_{i}$ in $S$ are scaled
by a factor $\alpha > 0$ the bound still holds. To see this one merely
has to observe that after scaling $\Pcal$ becomes $\alpha \Pcal$ and
$\Qcal$ becomes $\alpha \Qcal$, but the ratio $\Qcal/\Pcal$
which appears in \eqref{eq:additive-mult-practical} remains
unchanged. Thus while our algorithm is modeled as a generic
optimization formulation, it is possible to obtain scale invariance by
appropriately choosing $\epsilon'$ to be $\epsilon_{M}\Pcal^2$.


\section{Minimum Enclosing Convex Polytope}
\label{sec:MECP}

In the Minimum Enclosing Convex Polytope (MECP) problem we are given a
polytope of fixed shape which can be translated and magnified but
rotations are not allowed. Furthermore, we assume that the polytope has
a finite number of faces and hence it can be expressed as an
intersection of a finite number of hyperplanes:
\begin{align*}
  \inner{\wb_{i}}{\xb-\cbb} \leq t_{i} \qquad i= 1,2,....m,
\end{align*}
where $\cbb$ is the \emph{center} of the polytope about which it is
magnified. Given a set of points $S = \cbr{\xb_{1},\xb_{2},\hdots,\xb_{n}}$
we want to find the minimum magnification of the convex polytope that
encloses all the points in $S$.

\subsection{Formulation as an Optimization problem}
\label{sec:MECP_formulation} 

Clearly an enclosing polytope is one for which
$\inner{\wb_i}{\xb_j-\cbb} \leq t_i$ for all $i$ and $j$. This
observation helps us to cast the MECP problem as the following
optimization problem
\begin{align*}
  \min_{R \in \mathbb{R}, \cbb \in Q_1} & R \qquad \text{s.t. }  \inner
  {\wb_i}{\xb_j-\cbb} \leq R t_i
\end{align*}
or equivalently as 
\begin{align*}
  \min_{R \in \mathbb{R}, \cbb \in Q_1} & R \qquad \text{s.t. }
  \inner{\frac{\wb_i}{t_i}}{\xb_j-\cbb} \leq R \quad \forall i,j.
\end{align*}
Here $R$ is the scale of magnification of the polygon and
$Q_1\subset\RR^d$ is a bounded set, for example, a ball of certain fixed
radius $\Qcal$ centered at the origin which is assumed to contain the
solution $\cbb$. Usually $Q_{1}$ is taken to be a ball which contains
all the points in $S$ but this need not always be the case. Also we will
assume that $\wb_{i}/t_{i}$ lie inside a ball of radius $\Wcal$.
Writing $\tilde{\wb}_i = \wb_i/t_i$ the problem can be expressed as,
\begin{align}
  \label{eq:MECP_objective}
  \min_{\cbb \in Q_1}\max_{i,j}\inner{ \tilde{\wb}_i}{\xb_j-\cbb} =
  \min_{\cbb \in Q_1}\max_{\ub \in \Delta_{mn}} \sum_{i,j}u_{ij}(\inner{
    \tilde{\wb}_i}{\xb_j-\cbb}).
\end{align}
Notice that, as before, we have replaced the maximization over a
finite set by a maximization over the simplex. Clearly this problem
can be rewritten as
\begin{align}
  \label{eq:MECP_Nesterov}
  \min_{\cbb \in Q_1} J(\cbb) \text{ where } J(\cbb)  = \max_{\ub \in
    \Delta_{mn}} \sum_{i,j}u_{ij}(\inner{\tilde{\wb}_i}{\xb_j-\cbb}) 
   = \max_{\ub \in \Delta_{mn}}\cbr{\inner {\Ab\cbb}{\ub} + \inner{\ub}{\bb}}.
\end{align}
We used $\Ab = [\underbrace{ -\tilde{\wb}_1,\hdots}_{n \hspace{1 mm}
  \text{times}} \underbrace{-\tilde{\wb}_2,\hdots}_{n \hspace{1 mm}
  \text{times}}\hdots -\tilde{\wb}_m]^{\top}$, a $mn \times d$ matrix
with each $\tilde{\wb}_i$ repeated $n$ times as the columns and $\bb$ a
$mn$ dimensional vector with $b_{ij}=\inner{\wb_i}{\xb_j}$ to write the
above expression. Clearly $J(\cbb)$ can be identified with
\eqref{eq:primal} by setting $Q_2 = \Delta_{mn}$, $f(\cbb) = 0$ and
$g(\ub) = -\inner{\bb}{\ub}$. Here, $g$ is $0$-\lcg, however, $f$ is no
longer strongly convex. Therefore, we will work with the following
function
\begin{align*}
  J_{\eta}(\cbb) = \eta \nbr{\cbb}^{2} + \max_{\ub \in
    \Delta_{mn}}\cbr{\inner {\Ab\cbb}{\ub} - g(\ub)}.
\end{align*}
Let $J_{\eta}^{*} = \min_{\cbb \in Q_1} J_{\eta}(\cbb)$ and $J^{*} =
\min_{\cbb \in Q_1} J(\cbb)$. Since $\nbr{\cbb}^{2} \leq \Qcal^{2}$ for
all $c \in Q_1$ we have that
\begin{align*}
  J_{\eta}^{*} \leq J^{*} + \eta \Qcal^2.
\end{align*}
Suppose we can minimize $J_{\eta}$ to $\epsilon/2$ precision,
that is, find a $\cbb$ such that $J_{\eta}(\cbb) \leq J_{\eta}^{*} +
\epsilon/2$ then the above observation allows us to write the
following series of inequalities
\begin{align*}
  J(\cbb) \leq J_{\eta}(\cbb) \leq J_{\eta}^{*} + \frac{\epsilon}{2} \leq
  J^{*} + \frac{\epsilon}{2} + \eta \Qcal^{2}. 
\end{align*}
In other words, every $\epsilon/2$ accurate solution of
$J_{\eta}$ is a $\epsilon/2 + \eta \Qcal^{2}$ accurate solution of
$J$. In particular, if we set $\eta = \epsilon/2 \Qcal^{2}$ then
every $\epsilon/2$ accurate solution of $J_{\eta}$ is an
$\epsilon$ accurate solution of $J$. Furthermore, $J_{\eta}$ is nearly
identical to \eqref{eq:MEB-primal} except that $\nbr{\cbb}^{2}$ is now
replaced by $\frac{\epsilon}{2 \Qcal^{2}} \nbr{\cbb}^{2}$. Consequently,
Algorithm \ref{algo:MEB} can be directly applied to minimize $J_{\eta}$
with the following changes:
\begin{align}
  \label{eq:MECP-dual}
  D_{\eta}(\ub) & = \inner{\ub}{\bb} - \frac{\Qcal^{2}}{2 \epsilon} \ub^{\top}
  \Ab \Ab^{\top} \ub. \\
  \label{eq:MECP-dual-gradient}
  \nabla D_{\eta}(\ub) & = \bb - \frac{\Qcal^{2}}{\epsilon} \Ab \Ab^{\top}
  \ub. \\
  \nonumber V(\ub) &= \argmin_{\vb \in
    \Delta_{mn}}\cbr{\frac{L}{2}\|\vb-\ub\|^{2}
    - \inner{\grad D_{\eta}(\ub)} {\vb-\ub}} \\
  \label{eq:MECP:adj_gradient}
  &= \argmin_{\vb \in \Delta_{mn}}\cbr{\frac{L}{2}\|\vb-\ub\|^{2} -
    \inner{\bb - \frac{\Qcal^{2}}{\epsilon} \Ab \Ab^{\top} \ub} {\vb-\ub}}.
\end{align}
A simple application of the Cauchy-Schwartz inequality shows that 
\begin{align}
  \label{eq:MECP-grad-lipsch}
  \nbr{ \nabla D_{\eta}(\ub_{1}) - \nabla D_{\eta}(\ub_{2})} = \nbr{-
    \frac{\Qcal^{2}}{\epsilon} \Ab \Ab^{\top} \ub_{1} +
    \frac{\Qcal^{2}}{\epsilon} \Ab \Ab^{\top} \ub_{2}} \leq
  \frac{\Qcal^{2}}{\epsilon} \nbr{\Ab \Ab^{\top}} \nbr{\ub_{1} - \ub_{2}},
\end{align}
thus establishing that $D_{\eta}(\ub)$ is $\frac{\Qcal^{2}}{\epsilon}
\nbr{\Ab \Ab^{\top}}$-\lcg. We define $L_{\eta} =
\frac{\Qcal^{2}}{\epsilon} \nbr{\Ab \Ab^{\top}}$. Since
$\tilde{\wb_{i}}$ are assumed to lie inside a ball of radius $\Wcal$, by
an argument analogous to \eqref{eq:L-bound}, it follows that $L_{\eta} =
\Qcal^{2} \Wcal^{2}/\epsilon$. Plugging this into
\eqref{eq:convergence} obtains
\begin{align}
  \label{eq:MECP-convergence}
  J(\cbb_{k}) - D(\ub_{k}) \le \frac{3 \Qcal^{2} \Wcal^{2}}{\epsilon
    (k+1) (k+2)}.
\end{align}
In order to obtain an $\epsilon/2$ accurate solution we need to
solve for $k$ by setting $\frac{3 \Qcal^{2} \Wcal^{2}}{\epsilon (k+1)
  (k+2)} \leq \epsilon/2$. This yields $k \geq \sqrt{6}
\Qcal \Wcal/\epsilon$, which shows the $O(\Qcal/\epsilon)$ iteration
bound as claimed. Extending the arguments for MEB to the MECP case, the
per iteration complexity of $O(mnd)$ can readily be established. We omit
details for brevity.

\section{Applications to Machine Learning}
\label{sec:application}

The connection between the MEB problem and SVMs has been discussed in
a number of publications
\citep{Clarkson08,HarRotZim07,GaeJag09}. Practical algorithms using
coresets were also proposed in \citet{TsaKocKwo07} and
\citet{TsaKwoChe05}. In all these cases, our improved MEB algorithm
can be plugged in as a subroutine and will yield corresponding
speedups. We describe these kernel based algorithms and their
connection with MEB in appendix \ref{sec:kernels}. In this section, we
present two machine learning problems wherein our algorithms lead to
better bounds than existing coreset based approaches.

\subsection{Finding Large Margin Classifiers}
\label{sec:FindingLargeMargin}

In \citet{HarRotZim07} a coreset algorithm (Coreset SVM) for finding the
maximum margin hyperplane was described. It turns out that our MECP
algorithm can be specialized to their setting, and yields improved
bounds. Briefly, given $m$ labeled data points $(\zb_{i}, y_{i})$ with
$\zb_{i} \in \RR^{d}$ and $y_{i} \in \cbr{\pm 1}$ the maximum margin
hyperplane can be found by solving\footnote{\citet{HarRotZim07} use
  $\xb_{i}$ for the training data and $\wb$ for the hyperplane. Here we
  use $\zb_{i}$ and $\cbb$ respectively to be consistent with our
  notation.}
$\argmax_{\cbb \in \RR^{d}, \nbr{\cbb} = 1} \min_{i} y_{i}
  \inner{\zb_{i}}{\cbb}$. 
Equivalently, by defining $\tilde{\wb}_{i} = y_{i} \cdot \zb_{i}$ one
can solve
\begin{align}
  \label{eq:max-margin-eq}
  \argmin_{\cbb \in \RR^{d}, \nbr{\cbb} = 1} \max_{i}
  \inner{\tilde{\wb}_{i}}{-\cbb}.
\end{align}
The above problem can be identified with \eqref{eq:MECP_objective} if we
set $S$ to be the empty set and $Q_{1}$ to be $\cbr{\cbb \in \RR^{d}
  \text{ s.t. } \nbr{\cbb} = 1}$. With this substitution our MECP
algorithm can directly be applied to solve \eqref{eq:max-margin-eq}. In
this case $\Qcal = 1$ and the bound \eqref{eq:MECP-convergence} reduces
to
\begin{align}
  \label{eq:max-margin-bound}
  \frac{3 \Wcal^{2}}{\epsilon (k+1) (k+2)} \leq \frac{\epsilon}{2}.
\end{align}
Solving for $k$ shows that $\sqrt{6}\frac{\Wcal}{\epsilon}$ iterations
suffice to obtain an $\epsilon$ accurate solution of
\eqref{eq:max-margin-eq}. 

The Coreset SVM algorithm produces a multiplicative approximation. To
compare with the bounds given by \citet{HarRotZim07} we follow the same
scheme described in Section \ref{sec:MultversAddit}. Let $\rho^{*}$
denote the optimal margin, that is, $\min_{\cbb \in \RR^{d}, \nbr{\cbb}
  = 1} \max_{i} \inner{\tilde{\wb}_{i}}{-\cbb}$, and set $\epsilon =
\rho^{*} \epsilon_{M}$. Substituting into \eqref{eq:max-margin-bound} and
solving for $k$ shows that to produce a $\epsilon_{M}$ multiplicative
approximation with our algorithm
\begin{align}
  \label{eq:max-margin-mult}
  \sqrt{6} \rbr{\frac{\Wcal}{\rho^{*}}} \frac{1}{\epsilon_{M}} \leq k.
\end{align}
In contrast, \citet{HarRotZim07} compute a Coreset SVM $C$ of size  
\begin{align}
  \label{eq:peled-roth-bound}
  |C| = O\rbr{\rbr{\frac{\Wcal}{\rho^{*}}}^{2}\frac{1}{\epsilon_{M}}} 
\end{align}
in $|C|$ iterations. Furthermore, the computational complexity of
Coreset SVM is $O(md |C| + |C| T(|C|))$, where $T(|C|)$ is the cost of
training a SVM on $|C|$ points. Since each iteration requires only
$O(md)$ effort, our algorithm has an improved computational complexity
of $O\rbr{md \rbr{\frac{\Wcal}{\rho^{*}}}
  \frac{1}{\epsilon_{M}}}$. However, there is one notable difference
between the two algorithms. Coreset SVM produces a sparse solution in
terms of the number of support vectors, while our algorithm comes with
no such guarantees.

\subsection{Computing Polytope Distance}
\label{sec:CompPolytDist}

Given a set of points $S = \cbr{\xb_{1}, \xb_{2}, \ldots, \xb_{n}}$ the
polytope distance is defined as the shortest distance $\rho$ of any
point inside the convex hull of $S$, $\conv(S)$, to the origin
\citep{GaeJag09}. Equivalently, we are looking for the vector with the
smallest norm in $\conv(S)$. A variant is to compute the distance
between two polytopes given by the points $S_{+} = \cbr{\zb_{1},
  \zb_{2}, \ldots, \zb_{n}}$ and $S_{-} = \cbr{\zb_{1}', \zb_{2}',
  \ldots, \zb_{n'}'}$. This problem can be solved by finding the
polytope distance of the Minkowski difference of $\conv(S_{+})$ and
$\conv(S_{-})$ \citep{BenBre98}. Since the arguments for both cases are
by and large very similar we will stick with the simpler formulation in
this paper. The polytope distance problem has a number of applications
in machine learning. A partial (and by no means exhaustive) list of
relevant publications includes \citet{GaeJag09,BenBre98}, and
\citet{KeeSheBhaMur00}.

To analyze this problem in our setting we start with a technical
lemma. A version of this lemma also appears as Theorem A.2 in
\citet{BenBre98}.
\begin{lemma}
  The following two problems are duals of each other:
  \begin{align}
    \label{eq:polytope-primal}
    \min_{\cbb} J(\cbb) := \max_{i} \inner{\cbb - \xb_{i}}{\cbb} 
  \end{align}
  \begin{align}
    \label{eq:polytope-dual}
    \max_{\ub \in \Delta_{n}} D(\ub) := -\frac{1}{4} \ub^{\top} \Ab
    \Ab^{\top} \ub = -\frac{1}{4} \nbr{\Ab^{\top} \ub}^{2},
  \end{align}
  where $\Ab = -[\xb_{1},\xb_{2}, \ldots, \xb_{n}]^{\top}$. 
\end{lemma}
\begin{proof}
  First rewrite the objective function in \eqref{eq:polytope-primal} as 
  \begin{align*}
    J(\cbb) = \nbr{\cbb}^{2} + \max_{i} \inner{-\xb_{i}}{\cbb} =
    \nbr{\cbb}^{2} + \max_{\ub \in \Delta_{n}} \inner{\ub}{\Ab \cbb}.
  \end{align*}
  This can be identified with \eqref{eq:primal} by setting $Q_{1} =
  \RR^{d}$, $Q_{2} = \Delta_{n}$, $f(\cbb) = \nbr{\cbb}^{2}$, and
  $g(\ub) = 0$. The dual problem \eqref{eq:polytope-dual} can directly
  be read off from \eqref{eq:dual} by noting that that the Fenchel dual
  of $\nbr{\cdot}^{2}$ is $\frac{1}{4} \nbr{\cdot}^{2}$.
\end{proof}
Clearly \eqref{eq:polytope-dual} computes the vector with the smallest
norm in $\conv(S)$, which is equivalent to the polytope distance
problem. Furthermore, \eqref{eq:polytope-primal} and
\eqref{eq:polytope-dual} are identical to \eqref{eq:MEB-primal} and
\eqref{eq:MEB-dual} respectively with $\bb = \zero$. Therefore the
algorithm we described in Section \ref{sec:MEB} can be applied with
minor modifications to yield a $O(nd/\sqrt{\epsilon})$ algorithm for
this problem also. Since the \citet{GaeJag09} algorithm selects
coresets, at most $O(1/\epsilon)$ components of $\ub$ are
non-zero. However, in our case no such guarantees hold.

\section{Conclusions and Future Work}
\label{sec:Conclusions}

We presented a new approximation algorithm for the MEB problem whose
running time is $O(nd\Qcal/\sqrt{\epsilon})$. Unlike existing algorithms,
which rely heavily on geometric properties, our algorithm is motivated
and derived purely from a convex optimization viewpoint.  We extended
our analysis to the MECP problem and obtain a $O(mnd\Qcal/\epsilon)$
algorithm. Not only does our treatment yield an algorithm with better
bounds, but preliminary experimental results in Appendix~\ref{sec:exp}
suggest that our algorithm is competitive on problems with a large
number of data points.

A more general version of the MECP problem was studied by
\citet{Panigrahy04}.  In his setting the convex polytope is allowed
translations, magnifications, and rotations. A simple greedy approach
(very reminiscent of coresets) yields an (1+$\epsilon$) multiplicative
approximation algorithm which takes $O(1/\epsilon^2)$ iterations to
converge. Please consult \citet{Panigrahy04} for details.

A natural question to ask is the following: Can our algorithms be
extended to deal with arbitrary convex shapes? In other words, given an
arbitrary convex shape can we find the optimal magnification and
translation that is required to enclose the set of points
$\cbr{\xb_j}_{j=1}^n$ at hand. Unfortunately, a straightforward
extension seems rather difficult. Even though it is well known that
every convex shape can be described as an intersection of half planes,
the number of half planes need not be finite. In such a case our MECP
algorithm, which crucially relies on the number of half planes being
finite, is clearly not applicable. Somewhat surprisingly, we are able to
obtain a $O(1/\sqrt{\epsilon})$ algorithm for the MEB problem, even
though a ball is made up of an intersection of infinitely many half
planes. Clearly, the strong convexity of the objective function plays an
important role in this context. We are currently trying to characterize
such problems in the hope that this investigation will lead to efficient
algorithms for a number of other related problems.

Rotations are a natural concept when working with geometric
algorithms. This does not naturally carry over to our setting where we
use convex optimization. In order to introduce rotations one has to work
with orthogonal matrices, which significantly complicates the
optimization strategy. A fruitful pursuit would be to investigate if the
insights gained from coresets can be used to solve complicated
optimization problems which involve orthogonal matrices efficiently.
   
Even though we are only beginning to scratch the surface on exploring
connections between optimization and computational geometry, we firmly
believe that this cross pollination will lead to exciting new algorithms
in both areas.  


\newpage
 \bibliographystyle{plainnat}
\bibliography{SODA11}

\begin{thebibliography}{21}
\providecommand{\natexlab}[1]{#1}
\providecommand{\url}[1]{\texttt{#1}}
\expandafter\ifx\csname urlstyle\endcsname\relax
  \providecommand{\doi}[1]{doi: #1}\else
  \providecommand{\doi}{doi: \begingroup \urlstyle{rm}\Url}\fi

\bibitem[Badoiu and Clarkson(2002)]{BadCla02}
M.~Badoiu and K.L. Clarkson.
\newblock Optimal core-sets for balls.
\newblock In \emph{Computational Geometry: Theory and Applications}, 2002.

\bibitem[Bennett and Bredensteiner(1998)]{BenBre98}
K.~P. Bennett and E.~J. Bredensteiner.
\newblock Geometry in learning.
\newblock In C.~Gorini, E.~Hart, W.~Meyer, and T.~Phillips, editors,
  \emph{Geometry at Work}, Washington, D.C., 1998. Mathematical Association of
  America.
\newblock Available http://www.math.rpi.edu/{$\sim$}bennek/geometry2.ps.

\bibitem[Borwein and Lewis(2000)]{BorLew00}
J.~M. Borwein and A.~S. Lewis.
\newblock \emph{Convex Analysis and Nonlinear Optimization: Theory and
  Examples}.
\newblock CMS books in Mathematics. Canadian Mathematical Society, 2000.

\bibitem[Boyd and Vandenberghe(2004)]{BoyVan04}
S.~Boyd and L.~Vandenberghe.
\newblock \emph{Convex Optimization}.
\newblock Cambridge University Press, Cambridge, England, 2004.

\bibitem[Clarkson(2008)]{Clarkson08}
Kenneth~L. Clarkson.
\newblock Coresets, sparse greedy approximation, and the frank-wolfe algorithm.
\newblock In \emph{SODA '08: Proceedings of the nineteenth annual ACM-SIAM
  symposium on Discrete algorithms}, pages 922--931. Society for Industrial and
  Applied Mathematics, 2008.

\bibitem[Elzinga and Hearn(1972)]{ElzHea72}
D.~J. Elzinga and D.~W. Hearn.
\newblock The minimum covering sphere problem.
\newblock \emph{Management Science}, 19:\penalty0 96--104, 1972.

\bibitem[G\"{a}rtner and Jaggi(2009)]{GaeJag09}
Bernd G\"{a}rtner and Martin Jaggi.
\newblock Coresets for polytope distance.
\newblock In \emph{Annual Symposium on Computational Geometry}, 2009.

\bibitem[Har-Peled et~al.(2007)Har-Peled, Roth, and Zimak]{HarRotZim07}
Sariel Har-Peled, Dan Roth, and Dav Zimak.
\newblock Maximum margin coresets for active and noise tolerance learning.
\newblock In \emph{International Joint Conference on Artificial Intelligence},
  pages 836--841, 2007.

\bibitem[Hiriart-Urruty and Lemar\'echal(1993)]{HirLem93}
J.B. Hiriart-Urruty and C.~Lemar\'echal.
\newblock \emph{Convex Analysis and Minimization Algorithms, {I} and {II}},
  volume 305 and 306.
\newblock Springer-Verlag, 1993.

\bibitem[Keerthi et~al.(2000)Keerthi, Shevade, Bhattacharyya, and
  Murthy]{KeeSheBhaMur00}
S.S. Keerthi, S.~K. Shevade, C.~Bhattacharyya, and K.~R.~K. Murthy.
\newblock A fast iterative nearest point algorithm for support vector machine
  classifier design.
\newblock \emph{IEEE Transactions on Neural Networks}, 11\penalty0
  (1):\penalty0 124--136, January 2000.

\bibitem[Megiddo(1984)]{Meg84}
Nimrod Megiddo.
\newblock Linear programming in linear time when the dimension is fixed.
\newblock \emph{J. ACM}, 31\penalty0 (1):\penalty0 114--127, 1984.

\bibitem[Nesterov(2005{\natexlab{a}})]{Nesterov05}
Yurii Nesterov.
\newblock Smooth minimization of non-smooth functions.
\newblock \emph{Math. Program.}, 103\penalty0 (1):\penalty0 127--152,
  2005{\natexlab{a}}.

\bibitem[Nesterov(2005{\natexlab{b}})]{Nesterov05a}
Yurii Nesterov.
\newblock Excessive gap technique in nonsmooth convex minimization.
\newblock \emph{SIAM J. on Optimization}, 16\penalty0 (1):\penalty0 235--249,
  2005{\natexlab{b}}.
\newblock ISSN 1052-6234.

\bibitem[Nesterov(1983)]{Nesterov83}
Yurri Nesterov.
\newblock A method for unconstrained convex minimization problem with the rate
  of convergence {$O$}$(1/k^2)$.
\newblock \emph{Soviet Math. Docl.}, 269:\penalty0 543--547, 1983.

\bibitem[Panigrahy(2004)]{Panigrahy04}
Rina Panigrahy.
\newblock Minimum enclosing polytope in high dimensions.
\newblock \emph{CoRR}, cs.CG/0407020, 2004.

\bibitem[Pardalos and Kovoor(1990)]{ParKov90}
P.~M. Pardalos and N.~Kovoor.
\newblock An algorithm for singly constrained class of quadratic programs
  subject to upper and lower bounds.
\newblock \emph{Mathematical Programming}, 46:\penalty0 321--328, 1990.

\bibitem[Sch{\"o}lkopf and Smola(2002)]{SchSmo02}
B.~Sch{\"o}lkopf and A.~Smola.
\newblock \emph{Learning with Kernels}.
\newblock {MIT} Press, Cambridge, MA, 2002.

\bibitem[Tsang et~al.(2005)Tsang, Kwok, and Cheung]{TsaKwoChe05}
Ivor~W. Tsang, James~T. Kwok, and Pak-Ming Cheung.
\newblock Core vector machines: Fast svm training on very large data sets.
\newblock \emph{J. Mach. Learn. Res.}, 6:\penalty0 363--392, 2005.
\newblock ISSN 1532-4435.

\bibitem[Tsang et~al.(2007)Tsang, Kocsor, and Kwok]{TsaKocKwo07}
Ivor~W. Tsang, Andr{\'a}s Kocsor, and James~T. Kwok.
\newblock Simpler core vector machines with enclosing balls.
\newblock In \emph{Proc.\ Intl.\ Conf.\ Machine Learning}, pages 911--918,
  2007.

\bibitem[Welzl(1991)]{Welzl91}
Emo Welzl.
\newblock Minimum enclosing disks (balls and ellipsoids).
\newblock \emph{Lecture Notes in Computer Science}, 555:\penalty0 359--370,
  1991.

\bibitem[Yildirim(2008)]{Yildirim08}
E.~Alper Yildirim.
\newblock Two algorithms for the minimum enclosing ball problem.
\newblock In \emph{SIAM Journal on Optimization}, pages 1368--1391, 2008.

\end{thebibliography}

\newpage
\appendix
\section*{Appendix}

\section{Proof of theorem \ref{thm:excessive_condi_satisfy}}
\label{sec:app_proof_exc_cond_meet}

Our proof is by and large derived using results from
\citet{Nesterov05a}.  We begin with a technical lemma.
\begin{lemma}
\label{lemma:nesterov:helper_alpha}
(Lemma 7.2 of \citet{Nesterov05a})
  For any $\ub$ and $\ubbar$, we have
  \begin{align*}
    D(\ub) + \inner{\grad D(\ub)}{\ubbar - \ub} = \inner{\Ab \cbb(\ub) +
      \bb}{\ubbar} + \nbr{\cbb(\ub)}^{2}.
  \end{align*}
\end{lemma}
\begin{proof}
  Direct calculation by using \eqref{eq:MEB-dual},
  \eqref{eq:MEB-dual-gradient}, and \eqref{eq:c-min-cform} yields 
  \begin{align*}
    D(\ub) + \inner{\grad D(\ub)}{\ubbar - \ub} & = \inner{\ub}{\bb} -
    \frac{1}{4} \ub^{\top} \Ab \Ab^{\top} \ub +
    \inner{\bb - \frac{1}{2} \Ab \Ab^{\top} \ub}{\ubbar - \ub} \\
    & = \inner{\ubbar}{\bb} - \inner{\frac{1}{2} \Ab \Ab^{\top}
      \ub}{\ubbar} + \frac{1}{4} \ub^{\top} \Ab \Ab^{\top} \ub
    \\
    & = \inner{\Ab \cbb(\ub) + \bb}{\ubbar} + \nbr{\cbb(\ub)}^{2}.
  \end{align*}
\end{proof}
We first show that the initial $\wb_1$ and $\val_1$ satisfy the
excessive gap condition \eqref{eq:excessive_gap_condition}.  Since
$-D$ is $L$-\lcg (from \eqref{eq:MEB-grad-lipsch}), so
\begin{align*}
  D(\ub_{1}) & \ge D(\ub_{0}) + \inner{\grad D(\ub_{0})}{\ub_{1} -
    \ub_{0}} - \frac{L}{2} \nbr{\ub_{1} - \ub_{0}}^2 \\
  (\text{using defn. of } \ub_1 \text{ and } \eqref{eq:adj_gradient})& =
  \max_{\ub \in \Delta_n} \cbr{ D(\ub_0) + \inner{\grad
      D(\ub_0)}{\ub-\ub_0} -
    \frac{L}{2}\nbr{\ub-\ub_0}^2} \\
  (\text{using lemma } \ref{lemma:nesterov:helper_alpha}) & = \max_{\ub
    \in \Delta_n}\cbr{\inner{\ub}{\bb} + \inner{\Ab\cbb(\ub_0)}{\ub}
    + \nbr{\cbb(\ub_0)}^{2} - \frac{L}{2}\nbr{\ub-\ub_0}^2} \\
  (\text {using defn. of } d \text{ and } \mu_1) & = \max_{\ub \in
    \Delta_n}\cbr{\inner{\ub}{\bb} + \inner{\Ab\cbb(\ub_0)}{\ub}
    + \nbr{\cbb(\ub_0)}^{2} - \frac{\mu_1 \sigma}{2} \nbr{\ub - \ub_{0}}^{2}} \\
  (\text{using } \cbb_1 = \cbb(\ub_0)) & = \nbr{\cbb_1}^{2} + \max_{\ub
    \in
    \Delta_n} \cbr{\inner{\Ab\cbb_1}{\ub} + \inner{\ub}{\bb} -
    \frac{\mu_1 \sigma}{2} \nbr{\ub - \ub_{0}}^{2}} \\ 
  & \ge J_{\mu_1}(\cbb_1)
\end{align*}
which shows that our initialization indeed satisfies
\eqref{eq:excessive_gap_condition}. Second, we prove by induction that
the updates in Algorithm \ref{algo:MEB} maintain
\eqref{eq:excessive_gap_condition}.  We begin with two useful
observations. Using \eqref{eq:mu2-update} and the definition of
$\tau_{k}$, one can bound
\begin{align}
  \label{eq:tau-mu-bound}
  \mu_{k+1} = (1 - \tau_{k}) \mu_{k} = \frac{6}{(k+3)(k+2)}
  \frac{L}{\sigma} \geq \tau_{k}^{2} \frac{L}{\sigma}.
\end{align}
Let $\gammab := \ub_{\mu_{k}}(\cbb_{k})$. The optimality conditions for
\eqref{eq:u_max} imply $\inner{\mu_{k} \sigma (\gammab - \ub_{0}) - \Ab
  \cbb_{k} - \bb}{\ub - \gammab} \ge 0$ and hence
\begin{align}
  \label{eq:opt_cond_alpha}
  \mu_{k} \sigma \inner{\gammab - \ub_{0}}{\ub - \gammab} \ge \inner{\Ab
    \cbb_{k} + \bb}{\ub - \gammab}.
\end{align}
By using the update equation for $\cbb_{k+1}$ and the convexity of
$\nbr{\cdot}^{2}$ 
\begin{align*}
  J_{\mu_{k+1}}(\cbb_{k+1}) & = \nbr{\cbb_{k+1}}^{2} + \max_{\ub \in
    \Delta_n} \cbr{\inner{\Ab \cbb_{k+1}}{\ub} + \inner{\ub}{\bb} -
    \frac{\mu_{k+1} \sigma}{2} \nbr{\ub - \ub_{0}}^{2}} \\
  & = \nbr{(1-\tau_{k}) \cbb_{k} + \tau_{k} \cbb(\betab_k)}^{2}  \\
  & \; \; \; + \max_{\ub \in \Delta_n} \cbr{(1-\tau_{k}) \inner{\Ab
      \cbb_{k}}{\ub} + \tau_{k} \inner{\Ab \cbb(\betab_k)}{\ub} +
    \inner{\ub}{\bb} - (1-\tau_{k}) \frac{\mu_{k} \sigma}{2}
    \nbr{\ub - \ub_{0}}^{2} } \\
  & \le \, \, \max_{\ub \in \Delta_n} \left\{ (1-\tau_{k}) T_{1} +
    \tau_{k} T_{2} \right \},
\end{align*}
where 
\begin{align*}
  T_{1} &= \sbr{-\frac{\mu_{k} \sigma}{2} \nbr{\ub - \ub_{0}} +
    \inner{\Ab \cbb_{k} + \bb}{\ub} + \nbr{\cbb_{k}}^{2}} \qquad \text{and}\\
  T_{2} &= \sbr{\inner{\Ab \cbb(\betab_k) + \bb}{\ub} +
    \nbr{\cbb(\betab_k)}^{2}}.
\end{align*}
$T_{1}$ can be bounded as follows
\begin{align*}
  T_{1} & = - \frac{\mu_{k} \sigma}{2} \nbr{\ub - \ub_{0}}^{2} +
  \inner{\Ab \cbb_{k} + \bb}{\ub} + \nbr{\cbb_{k}}^{2} \\
  &= -\frac{\mu_{k}\sigma}{2}\nbr{\ub - \gammab}^2 - \frac{\mu_{k}
    \sigma}{2} \nbr{\gammab - \ub_{0}}^{2} - \mu_{k}
  \sigma \inner{\gammab - \ub_{0}}{\ub - \gammab} \\
  & \; \;\; + \inner{\Ab
    \cbb_{k} + \bb}{\ub} +  \nbr{\cbb_{k}}^{2} \\
  (\text{using } \eqref{eq:opt_cond_alpha}) & \leq -
  \frac{\mu_{k}\sigma}{2}\nbr{\ub - \gammab}^2 - \frac{\mu_{k}
    \sigma}{2} \nbr{\gammab - \ub_{0}}^{2} - \inner{\Ab \cbb_{k} +
    \bb}{\ub - \gammab} \\
  & \; \;\; + \inner{\Ab \cbb_{k} + \bb}{\ub} +  \nbr{\cbb_{k}}^{2} \\
  & = - \frac{\mu_{k}\sigma}{2}\nbr{\ub - \gammab}^2 - \frac{\mu_{k}
    \sigma}{2} \nbr{\gammab - \ub_{0}}^{2} + \inner{\Ab \cbb_{k}
    + \bb}{\gammab} +  \nbr{\cbb_{k}}^{2} \\
  (\text{using defn.\ of } \gammab) & = -\frac{\mu_{k}\sigma}{2}\nbr{\ub
    - \gammab}^2 +
  J_{\mu_{k}}(\wb_{k}) \\
  (\text{using induction assumption}) & \le -\frac{\mu_{k}
    \sigma}{2}\nbr{\ub - \gammab}^2 + D(\ub_{k}) \\
  (\text{using concavity of } D) & \le - \frac{\mu_{k}\sigma}{2}\nbr{\ub
    - \gammab}^2 + D(\betab_k) + \inner{\grad D(\betab_k)}{\ub_{k} -
    \betab_k},
\end{align*}
while $T_{2}$ can be simplified by using Lemma
\ref{lemma:nesterov:helper_alpha}:
\begin{align*}
  T_{2} = \inner{\ub}{\bb} + \inner{\Ab \cbb(\betab_k)}{\ub} +
  \nbr{\cbb(\betab_k)}^{2} = D(\betab_k) + \inner{\grad
    D(\betab_k)}{\ub - \betab_k}.
\end{align*}
Putting the upper bounds on $T_{1}$ and $T_{2}$ together, and using
\eqref{eq:tau-mu-bound} we obtain the following result.
\begin{align}
  \nonumber J_{\mu_{k+1}}(\cbb_{k+1}) & \le \max_{\ub \in \Delta_n}
  \left\{ (1-\tau_{k}) \sbr{-\frac{\mu_{k}\sigma}{2}\nbr{\ub -
        \gammab}^2 + D(\betab_k) +
      \inner{\grad D(\betab_k)}{\ub_{k} - \betab_k}} \right.\\
  \nonumber & \; \; \; + \left . \tau_{k}\sbr{D(\betab_k) + \inner{\grad
        D(\betab_k)}{\ub - \betab_k}} \right\} \\
  \label{eq:penultimate}
  & \le D(\betab_k) + \max_{\ub \in \Delta_n} \cbr{-\tau_{k}^{2}
    \frac{L}{2}\nbr{\ub - \gammab}^2 + \inner{\grad
      D(\betab_k)}{(1-\tau_{k})\ub_{k} + \tau_{k} \ub - \betab_k}}.
\end{align}
Let $\vb = (1-\tau_{k})\ub_{k} + \tau_{k} \ub$. By using the definition
of $\betab_k$ from Algorithm \ref{algo:MEB} observe that
\begin{align}
  \label{eq:into-v}
  (1-\tau_{k})\ub_{k} + \tau_{k} \ub - \betab_k = \tau_k (\ub -
  \gammab) = \vb - \betab_k. 
\end{align}
Furthermore, $\vb \in \Delta_{n}$ since it is a convex combination of
$\ub_{k} \in \Delta_{n}$ and $\ub \in \Delta_{n}$. Plugging
\eqref{eq:into-v} into \eqref{eq:penultimate}
\begin{align*} 
  J_{\mu_{k+1}}(\cbb_{k+1}) & \le D(\betab_k) + \max_{\vb \in \Delta_n}
  \cbr{-\frac{L}{2}\nbr{\vb - \betab_k}^2 + \inner{\grad
      D(\betab_k)}{\vb - \betab_k}} \\
  (\text{using } \eqref{eq:adj_gradient} \text{ and defn. of }\ub_{k+1})
  &= D(\betab_k) + \inner{\grad D(\betab_k)}{\ub_{k+1} - \betab_k} -
  \frac{L}{2}\nbr{\ub_{k+1} - \betab_k}^2 \\
  (\text{Since } -D \text{ is }L-\lcg) & \le D(\ub_{k+1}).
\end{align*}

\section{A linear time algorithm for a box constrained diagonal QP with 
  a single linear equality constraint}
\label{sec:simple_qp}
In this section, we focus on the following simple QP:
\begin{align}
  \label{eq:box_constrained_qp}
  \min \frac{1}{2} \sum_{i=1}^n &d_i^2 (\alpha_i - m_i)^2 \\
  \nonumber
  s.t. \qquad l_i \le &\alpha_i \le u_i \quad \forall i \in [n];  \\
  \nonumber
  \sum_{i=1}^n & \sigma_i \alpha_i = z.
\end{align}
Without loss of generality, we assume $l_i < u_i$ and $d_i \neq 0$ for all $i$.
Also assume $\sigma_i \neq 0$ because otherwise $\alpha_i$ can be solved independently.
  To make the feasible region nonempty, we also assume
\[
\sum_i \sigma_i (\delta(\sigma_i > 0) l_i + \delta(\sigma_i < 0) u_i)
\le z \le
\sum_i \sigma_i (\delta(\sigma_i > 0) u_i + \delta(\sigma_i < 0) l_i).
\]
The algorithm we describe below stems from \cite{ParKov90} and finds the exact optimal
solution in $O(n)$ time. 

With a simple change of variable $\beta_i = \sigma_i (\alpha_i - m_i)$, the problem is
simplified as
\begin{center}
\begin{minipage}[t]{6cm}
\begin{align*}
  \min \qquad \frac{1}{2} \sum_{i=1}^n & \dbar^{2}_i \beta_i^2 \\
  s.t. \qquad l'_i \le & \beta_i \le u'_i \quad \forall i \in [n];  \\
  \sum_{i=1}^n & \beta_i = z',
\end{align*}
\end{minipage}
\begin{minipage}[t]{1.2cm}
\vspace{4em}
where
\end{minipage}
\begin{minipage}[t]{6cm}
\begin{align*}
  l'_i &= \left\{ {\begin{array}{ll}
   \sigma_i (l_i - m_i) & \text{if } \sigma_i > 0  \\
   \sigma_i (u_i - m_i) & \text{if } \sigma_i < 0  \\
\end{array}} \right., \\
  u'_i &= \left\{ {\begin{array}{ll}
   \sigma_i (u_i - m_i) & \text{if } \sigma_i > 0  \\
   \sigma_i (l_i - m_i) & \text{if } \sigma_i < 0  \\
\end{array}} \right., \\
\dbar^2_i &= \frac{d_i^2}{\sigma_i^2}, \quad z' = z - \sum_i \sigma_i m_i.
\end{align*}
\end{minipage}
\end{center}

We derive its dual via the standard Lagrangian.
\begin{align*}
  L = \frac{1}{2} \sum_i \dbar^{2}_i \beta_i^2 - \sum_i \rho_i^+ (\beta_i
  - l'_i) + \sum_i \rho_i^- (\beta_i - u'_i) - \lambda \left(\sum_i \beta_i
  - z' \right).
\end{align*}
Taking derivative:
\begin{align}
\label{eq:dual_connect_simple_qp}
  \frac{\partial L}{\partial \beta_i} = \dbar^{2}_i \beta_i - \rho_i^+
  + \rho_i^- - \lambda = 0
  \quad \Rightarrow \quad
  \beta_i = \dbar^{-2}_i (\rho_i^+ - \rho_i^- + \lambda).
\end{align}
Substituting into $L$, we get the dual optimization problem
\begin{align*}
  \min D(\lambda, \rho_i^+, \rho_i^-) &= \frac{1}{2} \sum_i \dbar^{-2}_i
  (\rho_i^+ - \rho_i^- + \lambda)^2 - \sum_i \rho_i^+ l'_i + \sum_i
  \rho_i^+ u'_i - \lambda z' \\
  s.t. \qquad &\rho_i^+ \ge 0, \quad \rho_i^- \ge 0  \quad \forall i \in [n].
\end{align*}
Taking derivative of $D$ with respect to $\lambda$, we get:
\begin{align}
\label{eq:lambda_constraint_simple_qp}
  \sum_i \dbar^{-2}_i (\rho_i^+ - \rho_i^- + \lambda) - z' = 0.
\end{align}
The KKT condition gives:
\begin{subequations}
\label{eq:kkt_simple_qp}
\begin{align}
\label{eq:kkt_1_simple_qp}
  \rho_i^+ (\beta_i - l'_i) &= 0, \\
\label{eq:kkt_2_simple_qp}
  \rho_i^- (\beta_i - u'_i) &= 0.
\end{align}
\end{subequations}
Now we enumerate four cases.
\paragraph{1. $\rho_i^+ > 0$, $\rho_i^- > 0$.}  This implies that $l'_i =
\beta_i = u'_i$, which is contradictory to our assumption.
\paragraph{2. $\rho_i^+ = 0$, $\rho_i^- = 0$.}  Then by
\eqref{eq:dual_connect_simple_qp}, $\beta_i = \dbar^{-2}_i \lambda \in
      [l'_i, u'_i]$, hence $\lambda \in [\dbar^{2}_i l'_i, \dbar^{2}_i u'_i]$.
\paragraph{3. $\rho_i^+ > 0$, $\rho_i^- = 0$.}  Now by \eqref{eq:kkt_simple_qp}
and \eqref{eq:dual_connect_simple_qp}, we have $l'_i = \beta_i = \dbar^{-2}_i
(\rho_i^+ + \lambda) > \dbar^{-2}_i \lambda$, hence $\lambda < \dbar^{2}_i l'_i$
and $\rho_i^+ = \dbar^{2}_i l'_i - \lambda$.
\paragraph{4. $\rho_i^+ = 0$, $\rho_i^- > 0$.}  Now by \eqref{eq:kkt_simple_qp}
and \eqref{eq:dual_connect_simple_qp}, we have $u'_i = \beta_i = \dbar^{-2}_i
(-\rho_i^- + \lambda) < \dbar^{-2}_i \lambda$, hence $\lambda > \dbar^{2}_i u'_i$
and $\rho_i^- = -\dbar^{2}_i u'_i + \lambda$.\\

In sum, we have $\rho_i^+ = [\dbar^{2}_i l'_i - \lambda]_+$ and $\rho_i^- =
[\lambda - \dbar^{2}_i u'_i]_+$.  Now \eqref{eq:lambda_constraint_simple_qp} turns
into
\begin{align}
\label{eq:lambda_find_root}
  f(\lambda) := \sum_i \underbrace{\dbar^{-2}_i ([\dbar^{2}_i l'_i - \lambda]_+ -
    [\lambda - \dbar^{2}_i u'_i]_+ + \lambda)}_{=: h_i(\lambda)} - z' = 0.
\end{align}
In other words, we only need to find the root of $f(\lambda)$ in
\eqref{eq:lambda_find_root}.  $h_i(\lambda)$ is given by 
\begin{align}
\label{eq:hi_lambda}
h_i(\lambda) = \left\{
 \begin{array}{rl}
   l'_i & \hspace{2mm}\text{if }\hspace{2mm} \lambda < \dbar_i^{2}l'_i\\
   \lambda\dbar_i^{-2} & \hspace{2mm}\text{if }\hspace{2mm}
   \dbar_i^{2}l'_i \leq \lambda \leq \dbar_i^{2}u'_i\\
   u'_i & \hspace{2mm}\text{if }\hspace{2mm} \lambda > \dbar_i^{2}u'_i
 \end{array} \right.
\end{align}

Note that $h_i(\lambda)$ is a monotonically increasing function of $\lambda$, so the
whole $f(\lambda)$ is monotonically increasing in $\lambda$.  Since $f(\infty) \ge 0$
by $z' \le \sum_i u'_i$ and $f(-\infty) \le 0$ by $z' \ge \sum_i l'_i$, the root must
exist.  Considering that $f$ has at most $2n$ kinks (nonsmooth points) and is linear
between two adjacent kinks, the simplest idea is to sort $\cbr{\dbar^{2}_i l'_i,
\dbar^{2}_i u'_i : i \in [n]}$ into $s^{(1)} \le \ldots \le s^{(2n)}$.  If $f(s^{(i)})$
and $f(s^{(i+1)})$ have different signs, then the root must lie between them and can be
easily found because $f$ is linear in $[s^{(i)}, s^{(i+1)}]$.  This algorithm takes at
least $O(n \log n)$ time because of sorting.

However, this complexity can be reduced to $O(n)$ by making use of the fact that the
median of $n$ (unsorted) elements can be found in $O(n)$ time.  Notice that due to the
monotonicity of $f$, the median of a set $S$ gives exactly the median of function values,
\ie, $f(\MED(S)) = \MED(\cbr{f(x):x \in S})$.  Algorithm \ref{algo:linear_simple_qp}
sketches the idea of binary search.  The while loop terminates in $\log_2 (2n)$ iterations
because the set $S$ is halved in each iteration.  And in each iteration, the time
complexity is linear to $|S|$, the size of current $S$.  So the total complexity is $O(n)$.
Note the evaluation of $f(m)$ potentially involves summing up $n$ terms as in
\eqref{eq:lambda_find_root}.  However by some clever aggregation of slope and offset, this
can be reduced to $O(|S|)$.

\begin{algorithm}[t]
  \caption{$O(n)$ algorithm to find the root of $f(\lambda)$.
    Ignoring boundary condition checks. \label{algo:linear_simple_qp}}
  \KwIn{Function $f$}
  \KwOut{$\lambda^*$: Root of $f$}
  Initialize: {Set kink set $S \leftarrow \cbr{\dbar_i^2 l'_i : i \in [n]} \cup \cbr{\dbar_i^2
    u'_i: i \in [n]}$.}\;
  \While{$\abr{S} > 2$}{
    
    Find median of $S$: $m \leftarrow \MED(S)$.

        \If{$f(m) \ge 0$}{
          
          $S \leftarrow \cbr{x \in S : x \le m}$.
        }
        \Else {
          $S \leftarrow \cbr{x \in S : x \ge m}$.
        }
        
      }
      Return root $\frac{l f(u) - u f(l)}{f(u) - f(l)}$ where $S = \cbr{l,u}$.
\end{algorithm}

\section{Learning SVMs with MEB}
\label{sec:kernels}

Kernel methods in general and support vector machines (SVMs) in
particular have received significant recent research interest in machine
learning \citep{SchSmo02}. Underlying a SVM is a simple geometric
idea. Given a training set $\cbr{(\xb_{i}, y_{i})}_{i=1}^{n}$ of $n$
points $\xb_{i}$ labeled by $y_{i} \in \cbr{\pm 1}$ the aim is to find
the hyperplane which maximizes the margin of separation between points
from different classes. This is compactly written as the following
optimization problem (see \citet{SchSmo02} for details):
\begin{subequations}
  \label{eq:svm-primal}
  \begin{align}
    \min_{\wb \in \RR^d} \; \; & \; \; \frac{1}{2}\nbr{\wb}^2 + C
    \sum_{i=1}^{n} \xi_{i} \\
    \text{s.t.} \; \; & \; \; y_i(\inner{\wb}{\xb_i}+b)\geq 1 - \xi_{i}
    \text{ for all } i.
\end{align}  
\end{subequations}
Standard duality arguments (see \eg \citet{BoyVan04}) yield the
following dual Quadratic Programming (QP) problem
\begin{subequations}
  \label{eq:svm-dual}
  \begin{align}
    \max_{\val \in \RR^m} \; \; & \; \; \inner{\alphab}{\eb}
    -\frac{1}{2}\alphab^{\top}K\alphab \\ 
    \text{s.t.}  \; \; & \; \; \sum_{i=1}^{n} \alpha_{i} y_{i} = 0 \\
    &\; \; 0 \leq \alpha_{i} \leq C \text{ for all } i.
  \end{align}  
\end{subequations}
Here $K$ is a $m \times m$ matrix whose entries are given by $y_i y_j
\inner{\xb_i}{\xb_j}$ and $\eb$ denotes the vector of all ones. Since
the dual only depends on $\xb$ via the inner products
$\inner{\xb_i}{\xb_j}$ one can employ the kernel trick: map $\xi_{i}$
into a feature space via $\phi(\xb_{i})$ and compute the dot product in
the feature space by using a kernel function $k(\xb_{i}, \xb_{j}) :=
\inner{\phi(\xb_i)}{\phi(\xb_j)}$ \cite{SchSmo02}. The kernel trick
makes SVM rather powerful because simple linear decision boundaries in
feature space map into non-linear decision boundaries in the original
space where the datapoints live. 

A number of different techniques have been proposed for solving the
quadratic problem associated with SVMs. Of particular interest in our
context is the Core Vector Machine (CVM)
\citep{TsaKocKwo07,TsaKwoChe05}. The key idea of the CVM is the
observation that solving \eqref{eq:svm-dual} is equivalent to finding
the MEB 
of
the feature vectors.  The MEB problem in feature space can be written
as (see \cite{TsaKocKwo07} for details)
\begin{subequations}
  \label{eq:cvm-primal}
  \begin{align}
    \min_{\cbb \in \RR^d, R \in \RR} \; \; & \; \; R^2 \\
    & \; \; \nbr{\cbb - \phi(\xb_i)}^2 \leq R^2 \text{ for all } i.
  \end{align}
\end{subequations}
The dual of the above minimization problem then becomes (also see
\eqref{eq:MEB-dual})
\begin{subequations}
  \label{eq:cvm-dual}
  \begin{align}
    \max_{\val} \; \; & \; \; \sum_{i=1}^n \alpha_i K_{i,i} -
    \val^{\top} K \val \\
    \text{s.t. } \; \; & \; \; \alpha_i \geq 0, \qquad \sum_{i}\alpha_i
    = 1
  \end{align}
\end{subequations}
where $K_{i,j} = \inner{\phi(\xb_i)}{\phi(\xb_j)}$ as before. In
particular, if each $K_{i,i}$ equals a constant $c$ then it can be shown
that by a simple transformation that the standard SVM dual
\eqref{eq:svm-dual} and the CVM dual \eqref{eq:cvm-dual} can be
identified. Therefore, every iteration of the CVM algorithm
\citep{TsaKwoChe05} identifies an active set of points, and computes the
MEB of this active set. This is done via the coreset algorithm of
\cite{Panigrahy04}, hence the name core vector machine. In fact, our
algorithm for the MEB can directly be plugged into the CVM, and improves
the rates of convergence of the inner iteration from $O(nd/\epsilon)$ to
$O(nd/\sqrt{\epsilon})$. We hope that our algorithm with improved
convergence rates can also be used in similar machine learning
algorithms to speed up their convergence. 

\section{Experimental Results}
\label{sec:exp}

The aim of our experiments is to demonstrate the efficacy of
Algorithm~\ref{algo:MEB} and compare its performance with existing MEB
algorithms in terms of running time and number of iterations required
for convergence. Following \citep{Yildirim08} we generate data using a
random multivariate Gaussian distribution and vary $n$ the number of
data points and $d$ the dimensions. For each fixed $n$ and $d$ we
generate 5 random datasets and report the average performance of our
algorithm with the multiplicative guarantee $\epsilon = 10^{-3}$ in
Table \ref{tab:Table1}. Recall from Section \ref{sec:MultversAddit} that
the multiplicative guarantee $\epsilon$ is equivalent to the additive
tolerance $\epsilon\Pcal^2$, where $\Pcal$ is chosen via
\begin{align*}
  \Pcal = \frac{1}{2}\max_{\xb_i,\xb_j}\nbr{\xb_i- \xb_j}. 
\end{align*}

For reference we also reproduce the results reported in Table 1 of
\cite{Yildirim08}. Although not a fair comparison, the CPU times gives
an indication of the relative performance of various algorithms. In the
table BC refers to the coreset algorithm of \citet{BadCla02} while A1
and A2 are MEB algorithms implemented in \citet{Yildirim08}.

\begin{table} 
  \begin{tabular}{|cc|cccc|cccc|}
    \hline
    \multirow{2}{*}{n} & \multirow{2}{*}{d} & \multicolumn{4}{|c|}{Time} &
    \multicolumn{4}{|c|}{Iterations} \\ 
    & & {A1} & {A2} & {BC} & {Ours} & {A1} & {A2} & {BC} & {Ours} \\ 
    \hline
    500 & 10 & 0.06 & 0.03 & 0.12 & 0.04 & 168.7 & 44.5 & 435.5 & 44.2 \\
    \hline
    1000 & 10 & 0.15 & 0.03 & 0.14 & 0.10 & 330.7 & 41.6 & 344.4 & 54.5 \\
    \hline
    5000 & 20 & 1.7 & 0.36 & 3.11 & 1.08 & 246.8 & 46 & 464.2 & 69.7 \\
    \hline
    10000 & 20 & 4.46 & 0.58 & 4.65 & 3.40 & 319.2 & 36.3 & 334.4 & 105.2 \\
    \hline
    30000 & 30 & 27 & 6.45 & 24.59 & 10.43 & 446.4 & 103.6 & 409 & 77.8 \\
    \hline
    50000 & 50 & 71.62 & 16.87 & 68.78 & 18.69 & 429.8 & 98.4 & 415.1 & 54.5 \\
    \hline
    100000 & 100 & 287.99 & 77.74 & 268.11 & 83.18 & 451.7 & 119 & 422.6 & 63 \\
    \hline
  \end{tabular}
  \caption{Computational Results with $n>>d$ ($\epsilon = 10^{-3}$)}
  \label{tab:Table1}
\end{table}

Our algorithm performs significantly better than BC and A1, while being
comparable to A2. Furthermore, our algorithm usually takes smaller
number of iterations and particularly shines when the number of points
is large. Our implementation is preliminary, and we believe that our
algorithm will benefit from practical speed ups in much the same way
that algorithm A2 was obtained by improvising A1 to get rid of
redundancies.

\end{document}